\documentclass[final,5p,times,twocolumn,letterpaper]{elsarticle}

\usepackage{natbib}
\usepackage{cancel}
\usepackage{pifont} 
\usepackage{amsthm}         
\usepackage{amsmath} 	   
\usepackage{amssymb}       
\usepackage{amsfonts}
\usepackage{enumerate}
\usepackage{graphicx} 	    
\usepackage{verbatim}
\usepackage{epsfig,bbm}
\usepackage{color}
\usepackage{colortbl}
\usepackage{float}
\usepackage{multirow}
\usepackage{tensor}
\usepackage{enumitem}

\usepackage{subfig}

\definecolor{babyblue}{rgb}{0.54, 0.81, 0.94}
\definecolor{corn}{rgb}{0.98, 0.93, 0.36}

\begin{document}

\begin{frontmatter}

\title{Slow Contraction and the Weyl Curvature Hypothesis}

%PRD starts here
%\author{Anna Ijjas }
%\affiliation{Max Planck Institute for Gravitational Physics (Albert Einstein Institute), 30167 Hanover, Germany}
%\affiliation{Gottfried Wilhelm Leibniz Universit{\"a}t, 30167 Hanover, Germany}
%\author{Paul J. Steinhardt %\corref{cor1}
%}
%\email[]{steinh@princeton.edu (corresponding author)}
%\affiliation{Department of Physics, Princeton University, Princeton, NJ 08544, USA}

\author[add1]{Anna Ijjas }
\ead{ijjas@nyu.edu}

\address[add1]{Center for Cosmology and Particle Physics, Department of Physics, New York University, New York, NY 10003, USA}

\date{\today}

\begin{abstract}
Using the power of numerical relativity,  we show that, beginning from generic initial conditions that are far from flat, homogeneous and isotropic and have a large Weyl curvature, 
a period of slow contraction rapidly drives spacetime towards vanishingly small Weyl curvature as the total energy density grows, thus providing a dynamical mechanism that satisfies the  Weyl Curvature Hypothesis. We also demonstrate a tight correlation between the Weyl Curvature Hypothesis and ultralocal behavior for canonical scalar fields with a sufficiently steep negative potential energy density.
 \end{abstract}

%\maketitle


\end{frontmatter}

\section{Introduction}   
According to the Weyl Curvature Hypothesis~\cite{Penrose:1979azm}, in the vicinity of the cosmic singularity (and any resolution thereof),
spacetime must have a vanishingly small Weyl curvature in order to evolve towards the flat, homogeneous and isotropic (henceforth: {\it smooth}) geometry we observe today.
However, unless starting from a select, special set of initial conditions that all have small Weyl curvature, such a state appears to be difficult (if not impossible) to reach by way of a dynamical mechanism. Under generic initial conditions, extrapolating backwards in time in an expanding universe or extrapolating forwards in time in a contracting universe,  anisotropy is expected to dominate, leading to chaotic mixmaster behavior and an arbitrary large Weyl curvature~\cite{Penrose:1988mg}. 
 
In this letter, we demonstrate a resolution to this problem. Namely, we show that a period of slow contraction is a dynamical mechanism that ensures the Weyl Curvature Hypothesis is satisfied under a broad range of initial conditions including those that have a large Weyl curvature. 

{\it Slow contraction} \cite{Cook:2020oaj} is a scaling attractor solution of the generalized Friedmann equation,
\begin{equation}
\label{FRW-eq}
3H^2 = \frac{\rho_m^0}{a^3}+ \frac{\rho_r^0}{a^4} + \frac{\sigma^2}{a^6} - \frac{k^2}{a^2} + \frac{\rho_{\phi}^0}{a^{2\varepsilon}},
\end{equation}
in which the Friedmann-Robertson-Walker (FRW) scale factor $a$ decreases as a power of the proper FRW time $\tau$:
\begin{equation}
\label{scalingsol}
a(\tau) = \left(\frac{\tau}{\tau_0} \right)^{1/\varepsilon} \; {\rm with} \quad \varepsilon\gg3.
\end{equation}
Here, $H\equiv{\rm d}\,\ln a/{\rm d}\tau$ is the Hubble parameter,  
$\rho_m^0, \rho_r ^0, k$ and $\sigma$ are constants that characterize the individual contributions of pressureless matter, radiation, (zero-mode) spatial curvature and anisotropy to the total energy density at some initial time~$t_0$. The parameter $\rho_{\phi}^0$ denotes the initial energy density and
$\varepsilon$ denotes the equation of state sourced by the scalar field matter component $\phi$ that drives slow contraction. For a minimally-coupled canonical scalar that has a negative exponential potential $V(\phi)=-V_0e^{-\phi/M}$ as considered in this letter, 
\begin{equation}
\label{eos}
\varepsilon =1/(2M^2).
\end{equation}
Throughout, the scalar field $\phi$ is expressed in reduced Planck units, {\it i.e.}, $M_{\rm Pl}\equiv 8\pi G_{\rm N}\equiv1$ with $G_{\rm N}$ being Newton's constant, spacetime coordinates are expressed in units of the initial mean curvature $\Theta_0^{-1}$ and $V_0$ is expressed in mixed units of $M_{\rm Pl}^2\Theta_0^{-2}$. 

As originally imagined \cite{Khoury:2001wf,Erickson:2003zm}, for slow contraction to start, there had to exist a select, special patch that is nearly homogeneous and isotropic as described by Eq.~\eqref{FRW-eq}.
The small deviations from homogeneity and isotropy are quickly diluted, as the trajectories are being attracted towards the flat FRW solution $3H^2 \simeq\rho_{\phi}/a^{2\varepsilon} \propto 1/\tau^2$.
Then, after an extended period of slow contraction, the features of this one patch are extended over exponentially many Hubble scales simply because
physical lengths, {\it e.g.}, the distance between two black holes, decrease as $a \propto |\tau|^{1/\varepsilon}$ which is exponentially slower than the Hubble scale shrinks since $|H|^{-1}\propto a^{\varepsilon}\propto |\tau|$.

In an extensive set of studies~\cite{Cook:2020oaj,Ijjas:2020dws,Ijjas:2021gkf,Ijjas:2021wml,Kist:2022mew}, using the tools of mathematical and numerical relativity, it has been shown that there actually is no need to assume a special, select set of initial conditions for slow contraction to begin. Rather, smoothing  starts for a very broad range of initial conditions including those that lie far outside the perturbative regime of flat FRW geometries. This means, the solution in Eqs.~(\ref{scalingsol}-\ref{eos}) is an attractor of the full, non-linear Einstein-scalar field equations with a large basin of attraction.
This result is remarkable in that {\it to date} there exists no other dynamical smoothing mechanism for which robustness to initial conditions has been established in any comparable way. But the result is not sufficient to demonstrate that slow contraction satisfies the Weyl Curvature Hypothesis because robustness has only been established in the frame/gauge associated with the underlying numerical scheme. 

In this letter, employing the frame/gauge invariant diagnostic tools recently developed in Ref.~\cite{Ijjas:2023bhh}, we follow the evolution of curvature invariants -- the Weyl curvature and the Chern-Pontryagin invariant introduced below in Sec.~\ref{sec:defs} -- and show in Sec.~\ref{sec:evol} that they become vanishingly small. This demonstrates that the robustness of slow contraction to initial conditions does not depend on the frame or gauge choice. In Sec.~\ref{sec:ul}, we then establish ultralocality, {\it i.e.}, the fact that spatial gradients rapidly decay and become irrelevant during contraction, as a key to achieving robustness. 

The fact that slow contraction satisfies the Weyl Curvature Hypothesis lends new support for the idea that the big bang singularity might be resolved by means of a cosmological bounce that connects the slowly contracting phase with the subsequent expansion.
 
 \section{Curvature invariants}
 \label{sec:defs}
 
The {\it conformal Weyl curvature tensor} $\tensor{\cal C}{_\mu_\nu_\rho_\sigma}$ \cite{Weyl:1918pdp} (henceforth referred to as {\it Weyl tensor})  is 
the trace-free part of the Riemann tensor $\tensor{R}{_\mu_\nu_\sigma_\rho}$, {\it i.e.},
%weyl tensor definition
\begin{equation}
\label{Weyl-coo}
\tensor{\cal C}{_\mu_\nu_\rho_\sigma}
\equiv \tensor{R}{_\mu_\nu_\rho_\sigma}
+ {\textstyle \frac12}\left( \tensor{g}{_\mu_\nu_\rho_\zeta}\tensor{R}{^\zeta_\sigma} 
+ \tensor{g}{_\mu_\nu_\zeta_\sigma}\tensor{R}{^\zeta_\rho}\right)
 + {\textstyle \frac16}R \tensor{g}{_\mu_\nu_\rho_\sigma}.
\end{equation} 
Here, $g_{\mu\nu}$ is the spacetime metric, $\tensor{R}{_\mu_\nu}\equiv \tensor{R}{^\sigma_\mu_\sigma_\nu}$ is the Ricci tensor, $R\equiv \tensor{R}{^\mu_\mu}$ is the Ricci  scalar 
and $\tensor{g}{_\mu_\nu_\rho_\sigma} \equiv  g_{\mu\rho}g_{\nu\sigma} - g_{\mu\sigma}g_{\nu\rho}$.
Throughout, spacetime indices $(0-3)$ are denoted by Greek letters and spatial indices $(1-3)$ are denoted by Latin letters.  

The Weyl tensor  describes the `non-local' part of the gravitational field, {\it i.e.}, inhomogeneities and anisotropies of the spacetime geometry that are not sourced by a local stress-energy source. As its name suggests, $\tensor{\cal C}{_\mu_\nu_\rho_\sigma}$ is invariant under conformal transformations of the metric. 

It proves useful to introduce the (left) dual of the Weyl tensor,
\begin{equation}
\label{def-dualWeyl}
{}^{\ast}\tensor{\cal C}{_\mu_\nu_\rho_\sigma}
\equiv \frac12 \tensor{\chi}{_\mu_\nu^\tau^\zeta}\tensor{\cal C}{_\tau_\zeta_\rho_\sigma},
\end{equation}
where $ \tensor{\chi}{_\mu_\nu_\tau_\zeta}\equiv - \sqrt{|-g|}\tensor{\epsilon}{_\mu_\nu_\tau_\zeta}$ denotes the totally anti-symmetric Levi-Civita 4-form and $\tensor{\epsilon}{_\mu_\nu_\tau_\zeta}$ denotes the Levi-Civita tensor.

Combining the Weyl tensor and its dual, we obtain two curvature invariants:
\begin{equation}
\label{def-weylscalar}
{\cal C} \equiv \tensor{\cal C}{^\mu^\nu^\rho^\sigma}\tensor{\cal C}{_\mu_\nu_\rho_\sigma},
\end{equation}
called the {\it Weyl curvature}, and 
\begin{equation}
\label{def-CPscalar}
{\cal P} \equiv  {}^{\ast}\tensor{\cal C}{^\mu^\nu^\rho^\sigma}\tensor{\cal C}{_\mu_\nu_\rho_\sigma},
\end{equation}
called the {\it Chern-Pontryagin invariant}. 
As shown in Ref.~\cite{Ijjas:2023bhh}, the Weyl curvature ${\cal C}$ and the Chern-Pontryagin invariant ${\cal P}$ enable us to fully describe numerical relativity simulations of cosmological spacetimes in an invariant way and without losing information from the gauge/frame dependent scheme that underlies the simulation.

In 3+1 dimensional spacetimes, the Weyl tensor  is in general non-zero. If ${\cal C}$ and ${\cal P}$ are sufficiently small,  the corresponding classical spacetime is effectively {\it conformally equivalent} to a flat FRW  geometry. 
More exactly, throughout this letter, we define the `flat FRW' state as a state with $\bar{\cal C},\bar{\cal P}\lesssim {\cal O}(10^{-10})$ and (as is the case in all the examples discussed here) with matter minimally coupled in the Einstein frame. As detailed in Ref.~\cite{Ijjas:2023bhh}, since the invariants are of ${\cal O}({}^3R)^2$, this upper bound on ${\cal C}$ and ${\cal P}$  ensures that  spacetime is sufficiently close to flat FRW such that the leading order spatial curvature fluctuations are of quantum origin with an amplitude of ${\cal O}(10^{-4})$.
\\

In the remainder of this letter, we follow the evolution of the invariants ${\cal C}$ and ${\cal P}$ to demonstrate that slow contraction satisfies the Weyl Curvature Hypothesis.

 \section{Evolution of curvature invariants}
 \label{sec:evol} 
 
To simulate slowly contracting spacetimes, we numerically evolve the Einstein-scalar field equations,
\begin{align}
\label{E-eq1}
R_{\mu\nu} - {\textstyle \frac12} g_{\mu\nu}R &= \nabla_{\mu}\phi\nabla_{\nu}\phi -  g_{\mu\nu}\left( {\textstyle \frac12}\nabla_{\lambda}\phi\nabla^{\lambda}\phi - V_0e^{-\phi/M} \right),\\
\label{E-eq2}
\Box \phi &= (V_0/M)\,e^{-\phi/M},
\end{align}
using the same numerical relativity code as in Refs.~\cite{Cook:2020oaj,Ijjas:2020dws,Ijjas:2021gkf,Ijjas:2021wml,Kist:2022mew}. In the example presented below, $V_0=0.1$ and $M=0.19$.

The numerical scheme relies on a tetrad form of the field equations~(\ref{E-eq1}-\ref{E-eq2}) in which spacetime points are being represented by a set of four 4-vectors (tetrads). The tetrad vector components $\bar{E}_{\alpha}^{\mu}$ are supplemented by a set of geometric variables, the mean curvature $\Theta^{-1}$, the components of the shear tensor $\bar{\Sigma}_{ab}$ and the components of the spatial curvature tensor $\bar{N}_{ab}$ that describe how the tetrad deforms when moving from one point to another. 

A `bar' above each variable means that the variable is re-scaled by the appropriate  power of the mean curvature $\Theta^{-1}$ and is hence {\it dimensionless}. 
Evaluating the variables relative to the mean curvature ({\it i.e.}, in the homogeneous limit, the Hubble scale)  is essential to obtain the observationally relevant quantities. For example, the density parameters $\Omega_i \equiv \rho_i/(3\Theta^{-2})$ (see, {\it e.g.}, Eqs.~4.1-4.3 in Ref.~\cite{Ijjas:2020dws}) yield the observationally relevant quantities but the mere densities $\rho_i$ do not. 
In addition, using mean-curvature-normalized variables in our simulations enables us to follow the evolution of the entire initial Hubble volume that has a radius $\Theta_0$ during several hundred $e$-folds of contraction in $\Theta$.  (This is analogous to using `co-moving' coordinates as common in cosmology to follow an extended period of accelerated expansion in the scale factor without losing coordinate volume.)

%%FIGURE1
\begin{figure*}[!htb]
 \begin{center}
\includegraphics[width=5.00in,angle=-0]{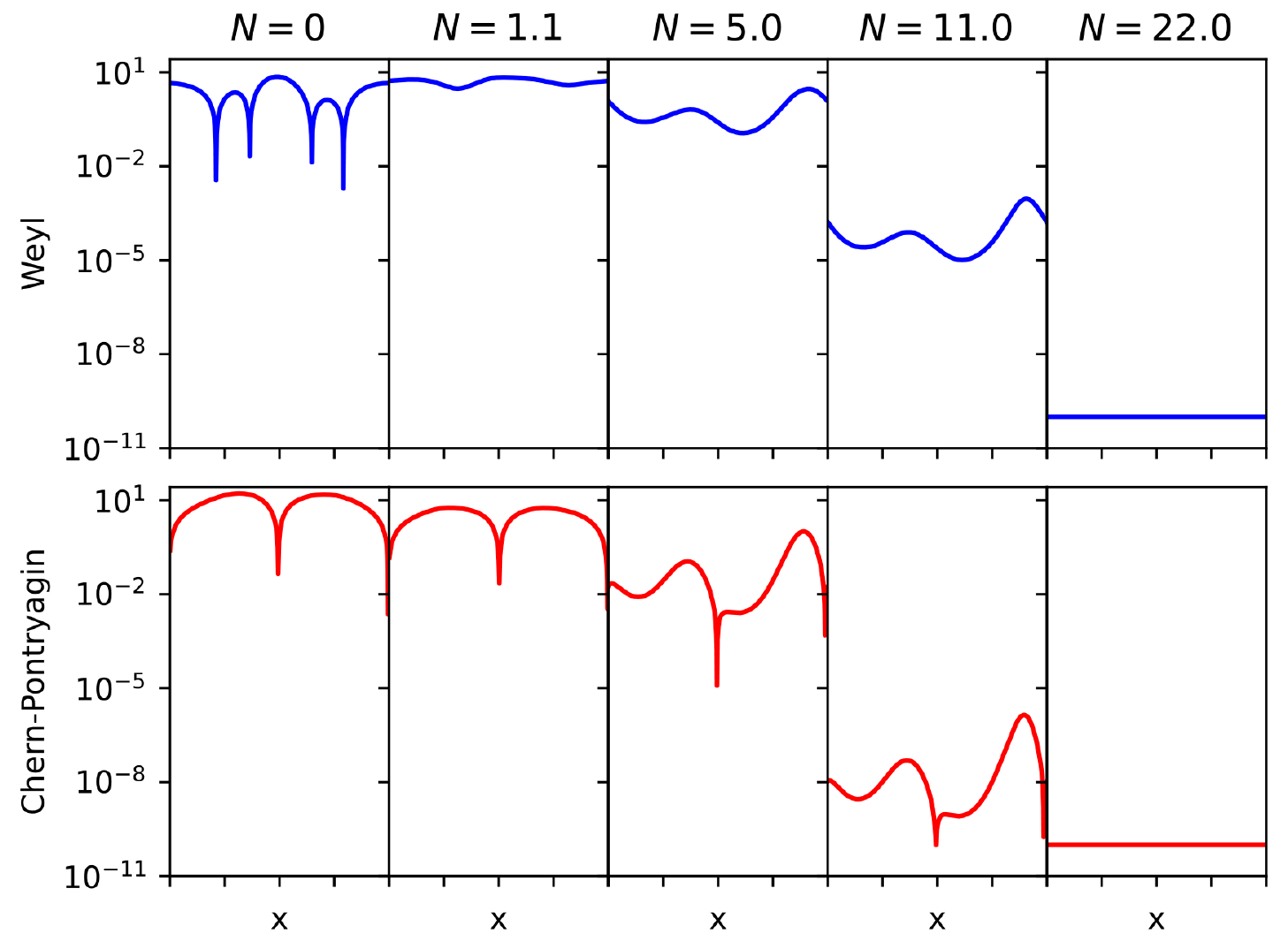}
\end{center}
\caption{Snapshots of the Weyl curvature $\bar{\cal C}$ (upper panel) and Chern-Pontryagin invariant  $\bar{\cal P}$ (lower panel) at sample $e$-fold times $N$ during slow contraction. The $x$ axis represents the entire simulation domain and the $y$ axis has dimensionless units by construction of the rescaled invariants. With our choice of coordinates, the simulation box size at time $N$ corresponds to a volume of radius $ 2\pi e^N \Theta_N$. For example, at $N=0$, the simulation box represents a volume with radius $2\pi\Theta_0$ (roughly, 6 Hubble volumes) but, after 22 $e$-folds of contraction, the same simulation box represents a volume with radius $ 2\pi e^{22}\Theta_N$ (roughly, $e^{22}$ Hubble volumes).  Complete smoothing ($\bar{\cal C}, \bar{\cal P}\lesssim {\cal O}(10^{-10})$) is reached everywhere by $N\sim22$.\label{Fig1}}
\end{figure*}  

We specify the initial conditions for our simulations by fixing the value of the geometric variables $\bar{E}_a^i,\Theta,\bar{\Sigma}_{ab}, \bar{N}_{ab}$ and the scalar field matter variables $\phi, \dot{\phi}$ at some initial time $t_0$.
Here, the only restriction is posed by the Hamiltonian and momentum constraints: In general, the field equations accept constraint violating initial data. But, if the constraints are satisfied at the initial time $t_0$, the field equations propagate them such that the constraints remain satisfied at all future times.
Employing the conformal York method \cite{York:1971hw}, as common in numerical relativity, we utilize all freedom left upon ensuring the Hamiltonian and momentum constraints are satisfied. In particular, we freely choose the conformally rescaled transverse, trace-free part of the initial shear tensor (also called `gravitational wave' contribution):
\begin{align}
\bar{\Sigma}_{ab}^0
= \psi^{-6}\Theta_0
\begin{pmatrix}
b_{2} & \xi &0 \\
\xi & b_{1}+a_{1} \cos x & a_{2} \cos x \\
0 &  a_{2} \cos x & -b_{1}-b_{2}-a_{1} \cos x
\end{pmatrix}
\label{eq:z_ab0}
\end{align}
 as well as the initial scalar field and velocity distributions:
 \begin{align}
\phi &= f_x\cos(n_x x+h_x) + \phi_0,
\label{eq:phi0} 
\\
\dot{\phi}
&= \psi^{-6}\Big(q_x\cos(m_x x+d_x) 
%+ q_y\cos(m_y y+d_y) 
+ Q_0\Big).
\label{eq:q0}
\end{align}
The conformal factor $\psi$ is determined by numerically solving the Hamiltonian constraint, as described, {\it e.g.}, in Ref.~\cite{Ijjas:2020dws}.
The sinusoidal form of the spatial variations is chosen to satisfy the periodic boundary conditions.

To reach the conclusions presented in this letter, we completed several hundred simulation runs, each lasting 120 or more  $e$-folds of contraction in the inverse mean curvature $\Theta$ and each corresponding to a different set of initial conditions ({\it i.e.}, a different combination of the parameters $\Theta_0, a_1, a_2, b_1, b_2, \xi, f_x, n_x, h_x, \phi_0, q_x, m_x, d_x, Q_0$) and/or a different choice of the potential $V(\phi)$. The example we discuss in the remainder of this letter is representative of our findings and corresponds to the following initial parameter values:
\begin{align}
\label{init1-1}
& \Theta_0 = 3;\; a_1,a_2=0.5,\; b_1=1.8,\; b_2=-0.15,\; \xi=0.01;
\\
& f_x = 0,\; n_x=1,\; h_x =-1.00,\; \phi_0=0;\\
& q_x =0.51,\; m_x=2,\; d_x=-1.7,\; Q_0 = 0.7.
\end{align}   

In Fig.~\ref{Fig1}, we show snapshots of the curvature invariants $\bar{\cal C}$ and $\bar{\cal P}$ at select times. 
At $N=0$, the value of both invariants is of ${\cal O}(10)$, proving that the initial conditions lie far outside the perturbative regime of flat FRW spacetimes. In only one $e$-fold, the system evolves away from the initial state and, in most regions, towards even larger values of $\bar{\cal C}$ and $\bar{\cal P}$.  This indicates that the initial state we chose is sufficiently generic. In particular, the initial state is not a stationary point or lies close to one.  By $N=5$, $\bar{\cal C}$ and $\bar{\cal P}$ are already starting to decrease. By $N=22$,  the flat FRW state is reached {\it everywhere} in the simulation domain.  For the subsequent $\sim 100$ $e$-folds of contraction, the system remains in the flat FRW state until the end of the simulation run with no sign of instability. Notably, the curvature invariants continue to shrink exponentially to become vanishingly small, satisfying the Weyl Curvature Hypothesis. At {\it e.g.}, $N=80$, $\bar{\cal C}\sim {\cal O}(10^{-50})$ and $\bar{\cal P}\sim {\cal O}(10^{-70})$.

It is not entirely surprising that complete smoothing is reached, given the earlier gauge/frame dependent analyses of Refs.~\cite{Cook:2020oaj,Ijjas:2020dws,Ijjas:2021gkf,Ijjas:2021wml,Kist:2022mew}. The result is nevertheless notable since there was no guarantee that the gauge/frame dependent description would carry over to an analysis in terms of curvature invariants. In fact, there are many historic examples (such as the evolution of primordial density fluctuations, see {\it e.g.} \cite{Bardeen:1980kt}) where the gauge/dependent description generally does {\it not} carry over to the calculation with gauge invariants.

\begin{figure}[!tb]
 \begin{center}
\includegraphics[width=0.9\linewidth,angle=-0]{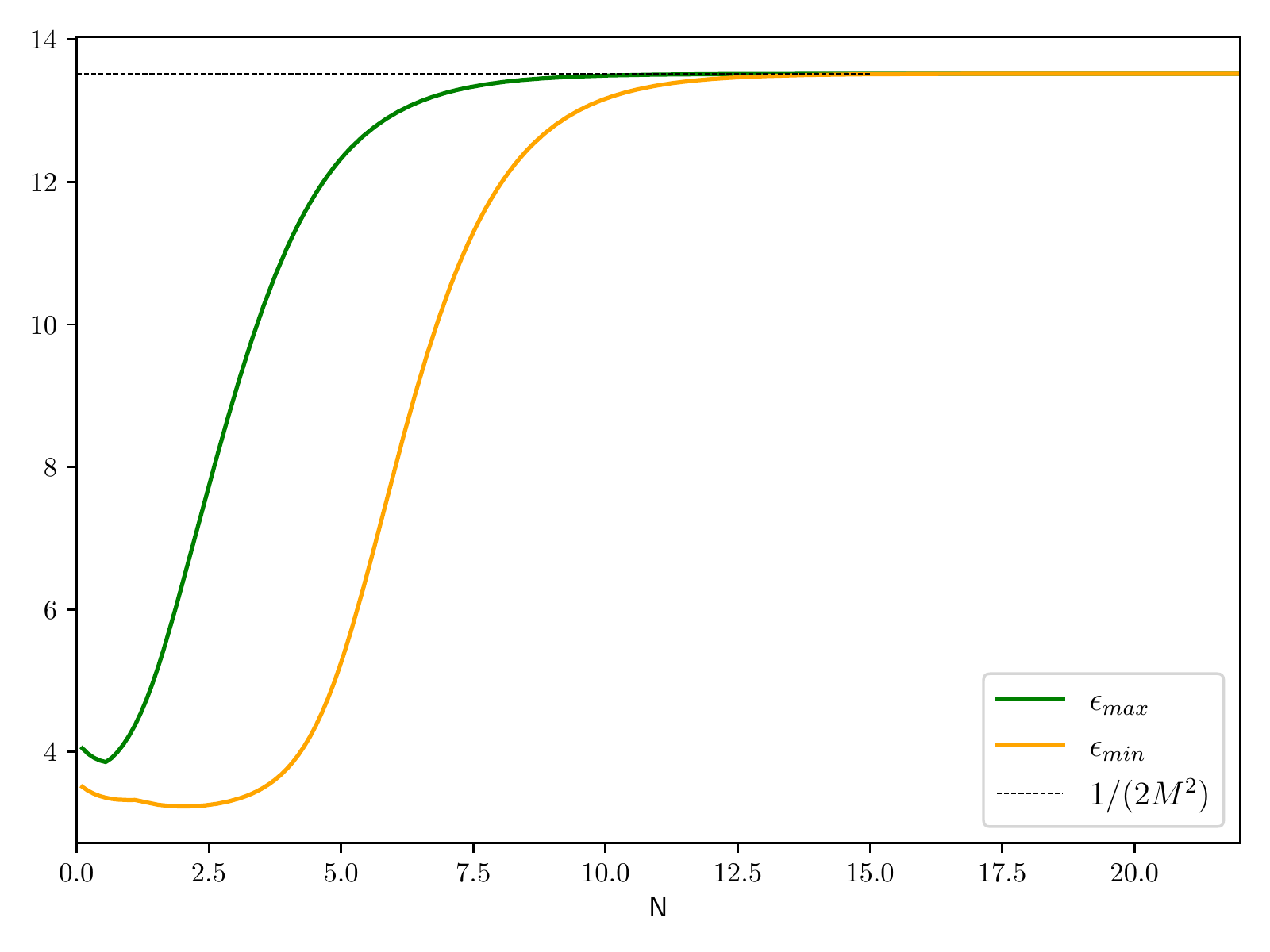}
\end{center}
\caption{Evolution of the maximum (green solid line) and minimum (orange solid line) of the equation of state parameter $\varepsilon$ as a function of $e$-fold time $N$. The first region reaches the scaling attractor solution given in Eq.~\eqref{scalingsol} with $\varepsilon=1/(2M^2)=13.85$ at $N\sim10$. By $N\sim15$, all regions have settled in the attractor solution. \label{Figeps}}
\end{figure}

\begin{figure}[!htb]
 \begin{center}
\includegraphics[width=0.9\linewidth,angle=-0]{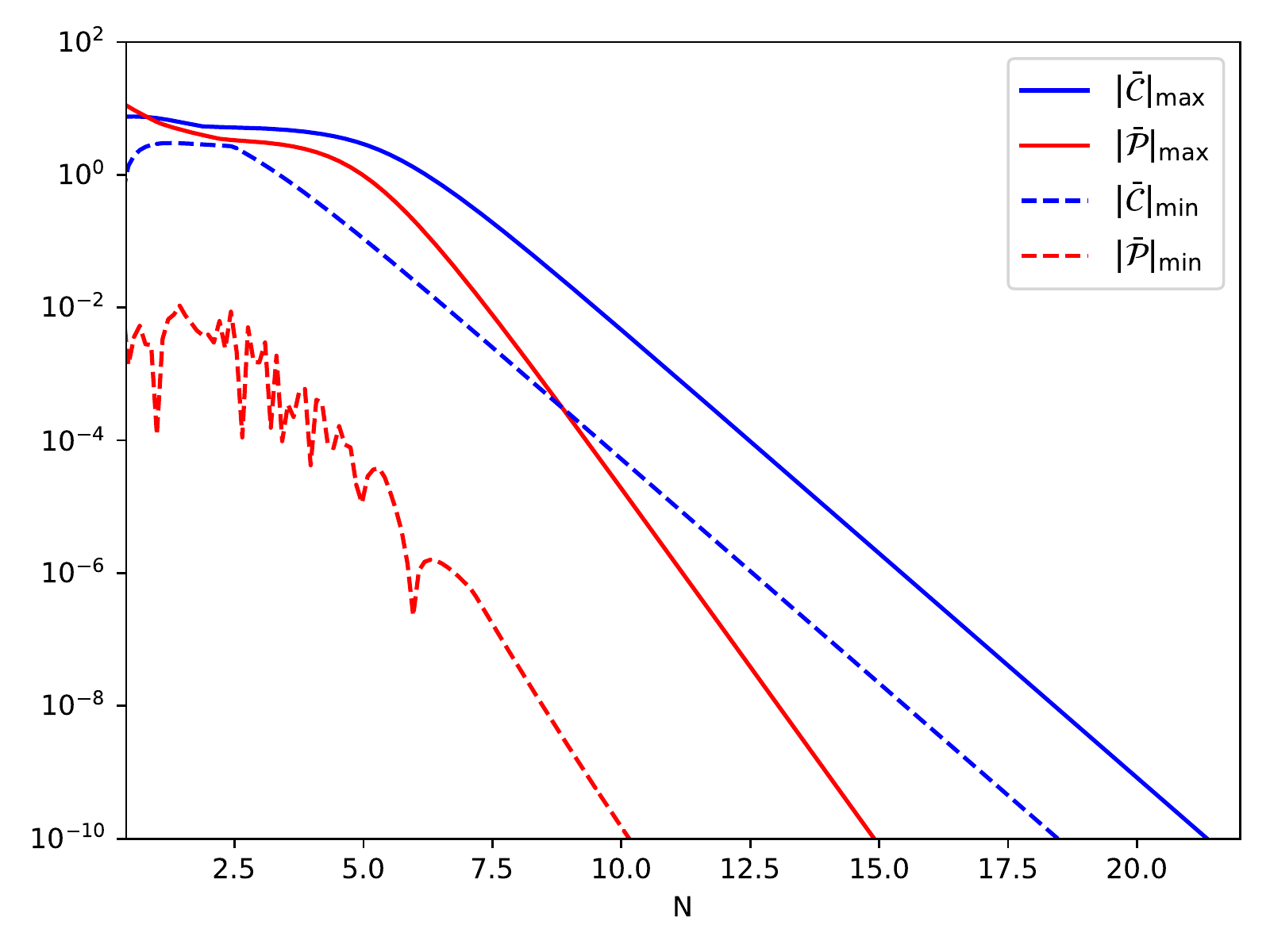}
\end{center}
\caption{(Log-scale) Evolution of the maximum (solid) and the minimum (dashed) values of the Weyl curvature $|\bar{\cal C}|$ (blue) and the Chern-Pontryagin invariant $|\bar{\cal P}|$ (red) as a function of $e$-fold time $N$. First, all four curves shrink only slowly but then decrease rapidly. \label{Fig2}}
\end{figure}

Intriguingly, even the details of the evolution are reminiscent of the gauge/frame dependent description:
\begin{enumerate}[leftmargin=0em]
\item[] First, it is apparent from Fig.~\ref{Fig1} that different spacetime points follow different trajectories, in agreement with the results of the gauge/frame dependent analysis of Refs.~\cite{Ijjas:2020dws,Ijjas:2021gkf,Ijjas:2021wml}.
\item[] Second, most of smoothing occurs well after spacetime split up into exponentially many causally disconnected (Hubble) volumes well before $N\sim15$ when all spacetime points reach the scaling attractor solution given in Eqs.~(\ref{scalingsol}-\ref{eos}) with $\varepsilon =1/(2M^2)$, as illustrated in Fig.~\ref{Figeps}. Here, $\varepsilon(x)\equiv1/{\cal N}(x)$, where ${\cal N}$ is the mean-curvature-normalized lapse function that, in the homogeneous limit, converges to the equation of state, as described, {\it e.g.}, in Ref.~\cite{Ijjas:2020dws}.
\item[] Third, the flat FRW state is reached in two  stages. At the onset of contraction, the curvature invariants $|\bar{\cal C}|$ and $|\bar{\cal P}|$  decrease only slowly ({\it i.e.}, by one or two orders of magnitude) but then after $\sim 5$ $e$-folds or so they start shrinking exponentially. This feature is already hinted in the snapshots of  Fig.~\ref{Fig1} but it becomes evident from Fig.~\ref{Fig2} where we show the evolution of the maximum (solid lines) and minimum (dashed lines) of the curvature invariants during the first 22 $e$-folds while the final flat FRW attractor state is being approached. 
\end{enumerate}

%%Figure2
%%

%
%%FIGURE3

\section{Ultralocality}
\label{sec:ul}

In previous numerical relativity studies of slowly contracting spacetimes using  gauge/frame dependent variables~\cite{Ijjas:2021gkf,Ijjas:2021wml}, the same evolution features  have been derived from {\it ultralocal} behavior. 
Here, a behavior is called `ultralocal' if spatial derivatives (henceforth: `gradients') in the equations of motion become small compared to the (first) time derivatives, see {\it e.g.} Ref.~\cite{Belinsky:1970ew}. Naively, one would {\it not} expect that a feature of the evolution equations that are frame and gauge dependent be reflected in a gauge/frame invariant description. However, in the following, we will demonstrate that this actually is the case.

\begin{figure}[!b]
 \begin{center}
\includegraphics[width=0.9\linewidth,angle=-0]{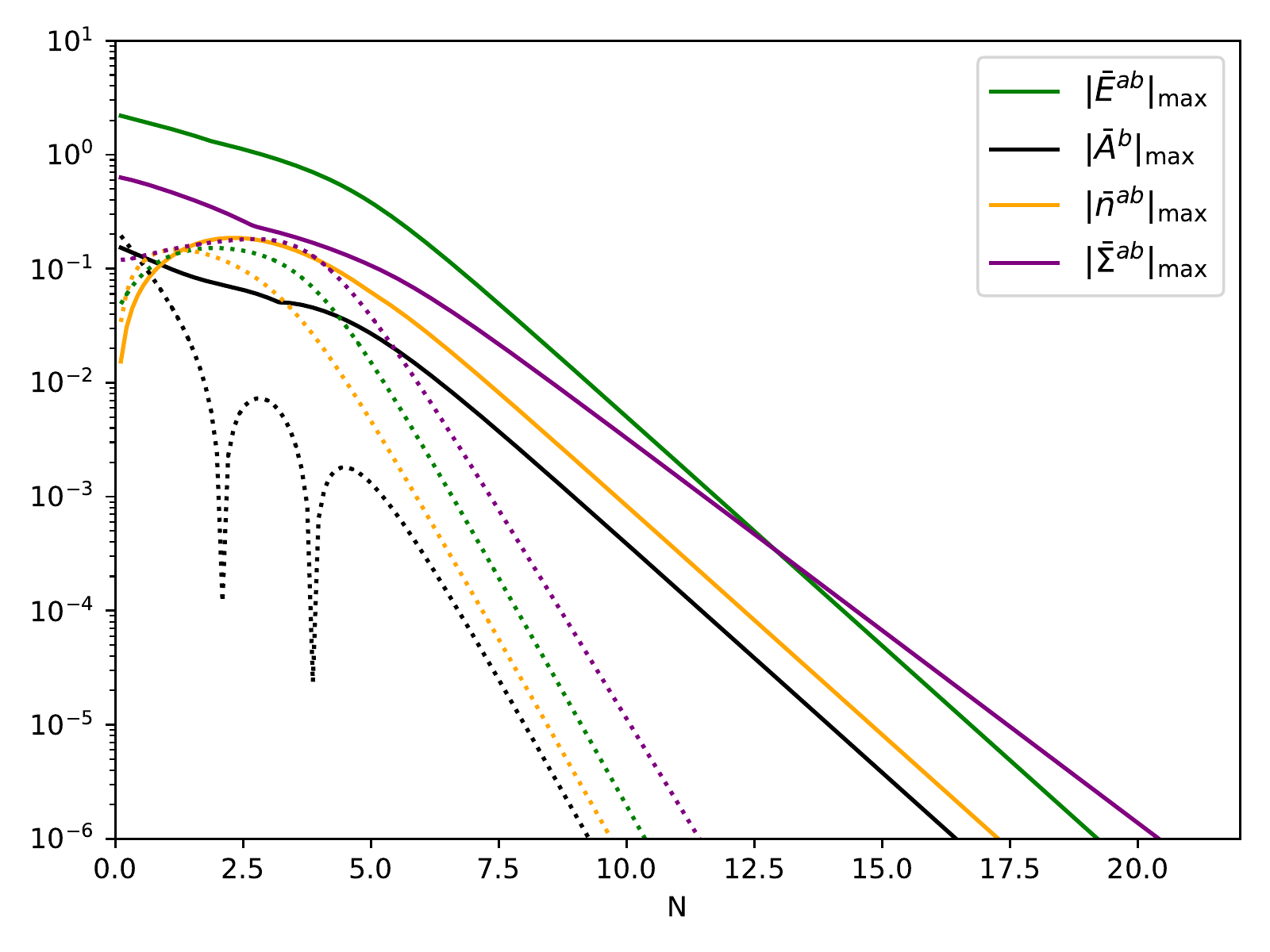}
\end{center}
\caption{Evolution of the maximum of the geometric variables (solid lines), $|\bar{E}_a^i|$ (green), $|\bar{A}_b|$ (purple), $|\bar{n}_{ab}|$ (orange), $|\bar{\Sigma}_{ab}|$ (black), and the maximum of the corresponding spatial gradients (dotted lines, same color coding) as a function of $e$-fold time $N$. \label{Figmax}}
\end{figure}

\begin{figure}[!hbt]
 \begin{center}
\includegraphics[width=0.9\linewidth,angle=-0]{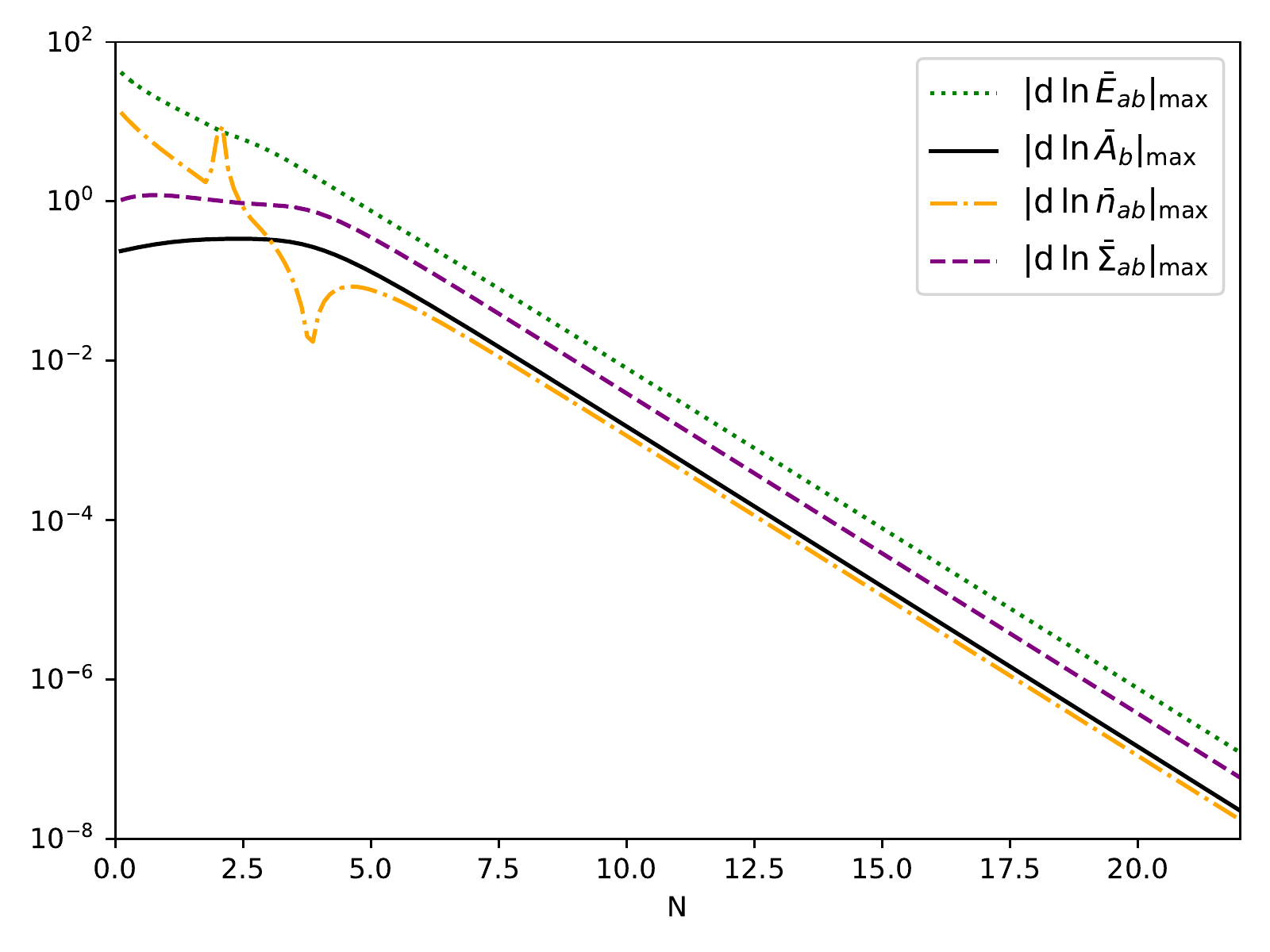}
\end{center}
\caption{The curves represent the maximum value of the ratios $|{\rm d}\bar{E}_a^i/\bar{E}_a^i|$ (green dotted), $|{\rm d}\bar{A}_b/\bar{A}_b|$ (purple dashed), $|{\rm d}\bar{n}_{ab}/\bar{n}_{ab}|$ (orange dashdot), $|{\rm d}\bar{\Sigma}_{ab}/\bar{\Sigma}_{ab}|$ (black solid) between as a function of $e$-fold time $N$. \label{Figdmax}}
\end{figure}

In Refs.~\cite{Ijjas:2021gkf,Ijjas:2021wml}, evidence for ultralocal behavior was found by analyzing the evolution of gauge/frame dependent dynamical variables for sample spatial points. With Figs.~\ref{Figmax}~and~\ref{Figdmax}, we give a complementary argument by showing characteristics of the entire system as it approaches the flat FRW attractor state. 

%%Figure4
%%Figure5
In Fig.~\ref{Figmax}, we show the maximum value of the geometric variables $|\bar{E}_a^i|$ (green), $|\bar{A}_b|$ (purple), $|\bar{n}_{ab}|$ (orange), $|\bar{\Sigma}_{ab}|$ (black) introduced in the previous section as well as the maximum of the corresponding gradients (dotted lines, same color coding). For $N\lesssim5$ all geometric variables (solid lines) and  the top three gradients (dotted lines) decrease only marginally. Then, at around $N\simeq5$ all gradients starts to shrink rapidly, reaching values of ${\cal O}(10^{-6})$ by $N\simeq11$. The geometric variables also start to decrease but this occurs first several $e$-folds later at $N\simeq7.5$. (NB: We also computed the minimum value of the geometric variables and, impressively, by $N\simeq22$ the maximum of all spatial gradients lie below the {\it minimum} of all geometric variables.) 

The fact that the maximum of all rescaled dimensionless gradients falls below of the value of all the rescaled, dimensionless geometric variables is a clear sign of ultralocality beginning by $N\simeq7.5$. 
As shown in Ref.~\cite{Ijjas:2020dws}, once the ultralocal limit is reached, the geometric variables all show the following simple scaling behaviors:
\begin{equation}
\bar{E}_a^i, \bar{A}_b, \bar{n}_{ab}\propto \Theta^{1-1/\varepsilon},\quad \bar{\Sigma}_{ab} \propto \Theta^{1-3/\varepsilon}.
\end{equation}
It is straightforward to verify these relation by simply inspecting the slope of the solid lines in Fig.~\ref{Figmax} after $N\simeq 7.5$.

Fig.~\ref{Figdmax} yields further evidence for ultralocal behavior. In the figure, we show the maximum of the ratio between the spatial gradients and the corresponding geometric variables, $|{\rm d}\bar{E}_a^i/\bar{E}_a^i|$ (green dotted), $|{\rm d}\bar{A}_b/\bar{A}_b|$ (purple dashed), $|{\rm d}\bar{n}_{ab}/\bar{n}_{ab}|$ (orange dashdot), $|{\rm d}\bar{\Sigma}_{ab}/\bar{\Sigma}_{ab}|$ (black solid), continuously shrinks. By $N\simeq7.5$, each ratio is smaller than one everywhere. Again, a sign of ultralocal behavior. By the time complete smoothing is reached at $N\simeq22$, the ratio is smaller than $10^{-6}$ everywhere.

Returning to Fig.~\ref{Fig2} and comparing the qualitative behavior of the curvature invariants to the frame/gauge dependent geometric variables in Figs.~\ref{Figmax}~and~\ref{Figdmax}, it is apparent that $|\bar{\cal C}|$ and $|\bar{\cal P}|$ start to shrink rapidly in every region of the simulation domain when the ultralocal state is reached at $N\simeq 7.5$, {\it i.e.}, the maximum of all spatial gradients is smaller than the maximum of all geometric variables and the maximum ratio between spatial gradients and the corresponding geometric variables (here: $|{\rm d}\bar{E}_a^i/\bar{E}_a^i|_{\rm max}$) is smaller than one. 

Quantitatively, using the formalism developed in Sec.~2 of Ref.~\cite{Ijjas:2023bhh}, we can identify the simple scaling solutions that the invariants obey in the ultralocal limit, namely
 \begin{equation}
\bar{\cal C} \propto \bar{\Sigma}^{ab}\bar{\Sigma}_{ab} \propto \Theta^{2(1-3/\varepsilon)},\quad
\bar{\cal P} \propto \bar{n}^{ab}\bar{\Sigma}_{ab} \propto \Theta^{2(1-2/\varepsilon)}.
\end{equation}
This explains the difference in the slopes of the two $\bar{\cal C}$ curves versus the two $\bar{\cal P}$ curves for $N>7.5$ in Fig.~\ref{Fig2}.

Based on our extensive numerical experiments combined with the analytic results in Ref.~\cite{Ijjas:2023bhh}, we conjecture that the ultralocal behavior of the underlying dynamical system shows itself through the simple scaling behavior of the curvature invariants.

\section{\bf Discussion}

The Weyl Curvature Hypothesis ties the problem of resolving the cosmic singularity with solving the cosmic initial conditions problem. To date, cosmologies that involve a big bang or quantum gravity beginning have not been shown to satisfy the hypothesis. The problem is that the proposed conditions are the opposite of what is expected after a big bang and a period dominated by large quantum gravity fluctuations.

In this letter, using numerical relativity, we have demonstrated that slow contraction satisfies the Weyl Curvature Hypothesis by driving spacetime towards a flat FRW attractor state with vanishingly small values of the Weyl curvature and the Chern-Pontryagin invariant {\it everywhere} in the simulation domain when starting from generic initial conditions. This result strengthens the argument for pursuing approaches that replace the big bang with a big bounce.

The key to {\it universal} smoothing as well as to reaching a vanishingly small Weyl curvature everywhere is the result of the interplay of two features unique to slow contraction:
the ultralocal behavior of contracting spacetimes combined with the fact that, in contracting spacetimes, the flat FRW solution has a large basin of attraction if the stress-energy is sourced by a minimally coupled canonical scalar field with a sufficiently steep negative potential energy density~\cite{Ijjas:2021gkf,Ijjas:2021wml}.

 \vspace{0.1in}
\noindent
{\it Acknowledgements.} 
Many thanks to  Roger Penrose, Paul J. Steinhardt, V. (Slava) Mukhanov, David Garfinkle, Brian Keating and Frans Pretorius for useful discussions.
This work is supported by the Simons Foundation grant number 947319.

\bibliographystyle{apsrev}
\bibliography{weyl}

\end{document}